\documentclass[a4paper,onecolumn,noarxiv]{quantumarticle}

\usepackage[utf8]{inputenc}
\usepackage[T1]{fontenc}
\usepackage{graphicx}
\usepackage{amsmath,amssymb}
\usepackage{bm}
\usepackage{booktabs}
\usepackage{xcolor}
\usepackage[colorlinks=true,linkcolor=blue,citecolor=blue,urlcolor=blue]{hyperref}

\graphicspath{{figures/}}

\begin{document}

\title{A four-player potential game for barren-plateau-aware quantum ansatz design}

\author{Ruben Dario Guerrero}
\email{rudaguerman@gmail.com}
\affiliation{Parametrized-QC-Graphs Project}

\date{\today}

\begin{abstract}
We cast the design of parameterized quantum circuits as a four-player potential game whose state is a circuit directed acyclic graph (DAG) and whose players encode trainability, non-stabilizerness, task performance, and hardware cost. Per-player restricted action sets factorize the move space into append, remove, retype, and rewire operations; a block-coordinate $\varepsilon$-Nash residual $\delta_\text{Nash}$ certifies that no single player can improve unilaterally. A single weight sweep on MaxCut $K_4$ traces a Pareto frontier from a Clifford endpoint $(M_2/n,\langle H\rangle)=(0,4.00)$ to a non-Clifford endpoint $(0.48,3.30)$. On three four-qubit hardware topologies (heavy-hex, $2\times 2$ grid, Rydberg all-to-all), Nash search achieves the highest mean potential; on the $2\times 2$ grid Nash reaches the theoretical ceiling $\Phi_\text{max}=4.10$ on two of five seeds while the simulated-annealing baseline does so on one; paired Wilcoxon tests over five seeds cannot reject the null on any single topology ($p\ge 0.22$). On LiH/STO-3G, seeding Nash from a 58-gate Givens-doubles ansatz produces a 48-operation, depth-25 circuit retaining $97.7\%$ of the correlation energy while simultaneously reducing gate count, increasing non-stabilizerness, and controlling trainability. The framework is complementary to energy-only searches such as ADAPT-VQE and k-UpCCGSD, which reach chemical accuracy with fewer operations but do not optimize the other three axes.
\end{abstract}

\maketitle

\section{Introduction}
\label{sec:intro}

Variational quantum algorithms (VQAs) offload the preparation of an entangled state to a parameterized quantum circuit (PQC) trained by a classical outer loop \cite{cerezo2021variational,bharti2022noisy}. Two complementary obstacles threaten their scaling. Barren plateaus cause the variance of every loss gradient to decay exponentially in qubit count whenever the dynamical Lie algebra (DLA) is exponentially large \cite{mcclean2018barren,ragone2024lie}; at the other extreme, classical simulation techniques that exploit a small DLA or a low stabilizer R\'enyi entropy render many barren-plateau-free circuits tractable on classical hardware \cite{cerezo2025does,bravyi2016trading,leone2022stabilizer}. Cerezo \textit{et al.} \cite{cerezo2025does} framed this as a structural tension: provably trainable VQAs in the literature are also provably simulable, so a convincing quantum advantage requires circuits placed in the interior of a trade-off space, not at either extreme.

Existing architecture searches attack the problem from a single direction. Hardware-efficient ans\"atze \cite{kandala2017hea} maximize expressibility at the cost of trainability. ADAPT-VQE \cite{grimsley2019adapt} grows operator pools adapted to chemistry and routinely reaches chemical accuracy on LiH with $O(20)$ operations. Differentiable and stochastic architecture searches such as DQAS \cite{zhang2022dqas} score candidate structures by task loss alone. None of these methods exposes the trainability/simulability trade-off as a controllable axis; none returns a stopping criterion that certifies balance among competing objectives.

In this work we (i) formalize PQC architecture search as a four-player potential game over circuit DAGs in which each player owns a restricted action set on the shared structure, (ii) demonstrate that a single weight sweep traces the Pareto frontier of the barren-plateau/simulability tension on MaxCut $K_4$, and (iii) show that the framework refines a chemistry-aware LiH ansatz while jointly controlling trainability, non-stabilizerness, and hardware cost. We report all benchmarks with honest uncertainty: our five-seed paired tests against a simulated-annealing baseline cannot reject the null on any single hardware topology despite a positive mean advantage, and we acknowledge that ADAPT-VQE reaches lower gate count than Nash at chemical accuracy on LiH. The contribution is orthogonal: Nash navigates a trade-off that energy-only methods do not see.

\section{Framework}
\label{sec:framework}

\subsection{Circuit DAG as a generalized graph-state adjacency}

A PQC is encoded as a directed acyclic graph $D=(V_\text{op}\cup V_\text{io},E_d)$ whose operation nodes $V_\text{op}=\{g_1,\ldots,g_L\}$ carry a gate type from a native set $\mathcal G$ and continuous parameters $\bm\theta$, and whose edges carry qubit-wire labels. This DAG is a strict generalization of the graph-state adjacency matrix \cite{hein2004graphstates,raussendorf2001oneway}: a graph state $|G\rangle$ lifts into a DAG with only CZ operations, while a generic PQC DAG allows heterogeneous gate types, continuous parameters, and multi-layer temporal ordering. The lowering $D\mapsto U(\bm\theta)$ is unique up to commuting-gate reordering and is implemented in a single codepath via TensorCircuit \cite{zhang2023tensorcircuit} with JAX JIT on an NVIDIA RTX\,4060 GPU.

\subsection{Players, actions, and potential}

We define four payoffs on the pair $(D,\bm\theta)$:
\begin{align}
f_1 &= d_\text{eff}(D,\bm\theta) & &\text{(trainability)}\nonumber\\
f_2 &= M_2(|\psi_{D,\bm\theta}\rangle)/n & &\text{(non-stabilizerness)}\label{eq:players}\\
f_3 &= -\langle H\rangle_{D,\bm\theta}\text{ or }+\langle H\rangle & &\text{(task)}\nonumber\\
f_4 &= C_\text{hw}(D) & &\text{(hardware cost)}\nonumber
\end{align}
Here $d_\text{eff}$ is derived from the quantum Fisher information matrix spectrum and controls gradient variance through the DLA bound $\mathrm{Var}[\partial_l\mathcal L]\in\Theta(1/\dim\mathfrak g)$ \cite{ragone2024lie}. The stabilizer R\'enyi-$2$ entropy $M_2$ \cite{leone2022stabilizer} is a practical non-stabilizerness proxy whose magnitude lower-bounds the Bravyi--Gosset simulation cost \cite{bravyi2016trading}. The task term $f_3$ is the minimized Hamiltonian expectation for VQE tasks and the maximized cut value for MaxCut. The hardware cost $C_\text{hw}$ counts native gates under the target connectivity.

Crucially, each player $i$ owns a \emph{restricted action set} $\mathcal A_i$ that only player $i$ may exercise on the shared DAG. Player $f_1$ (trainability) can retype a gate into one that enlarges the DLA; player $f_2$ (non-stabilizerness) can retype into non-Clifford primitives; player $f_3$ (task) can rewire two-qubit connections; player $f_4$ (hardware) can remove a gate. Every structural move is restricted to respect the target hardware's connectivity graph.

The potential
\begin{equation}
\Phi(D,\bm\theta)=w_1 f_1+w_2 f_2+w_3 f_3-w_4 f_4
\label{eq:potential}
\end{equation}
is a weighted scalarization. We emphasize that the potential form itself is unremarkable: any weighted scalarization of a multi-objective optimization is a potential game in the sense of Monderer and Shapley \cite{monderer1996potential}. The non-trivial content lies in the \emph{factorization of moves} into per-player restricted action sets, which gives each objective its own gradient-descent direction in circuit-structure space and allows a meaningful residual test.

\subsection{Nash certificate}

For each player $i$, the restricted best-response gap
\begin{equation}
\delta^{(i)}_\text{Nash}=\max_{a\in\mathcal A_i}\bigl[f_i(a\cdot D,\bm\theta)-f_i(D,\bm\theta)\bigr]_+
\label{eq:delta}
\end{equation}
measures how much player $i$ could gain by a unilateral deviation from $(D,\bm\theta)$. A point is an $\varepsilon$-Nash equilibrium when $\delta_\text{Nash}\equiv\max_i\delta^{(i)}_\text{Nash}\le\varepsilon$. This is the standard coordinate-wise $\varepsilon$-Nash residual adapted to the restricted-action per-player move sets; viewed from the optimization side it is a block-coordinate stationarity condition. It provides a sharper stopping criterion than the single-objective loss plateau used in DQAS and in simulated-annealing baselines.

\subsection{Optimizer}

\begin{figure*}[!tbp]
\centering
\includegraphics[width=\textwidth]{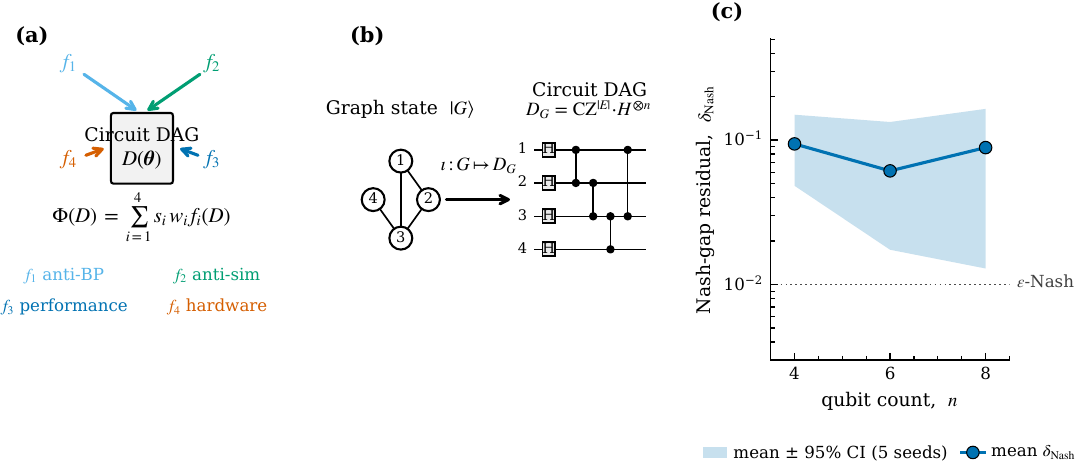}
\caption{\textbf{Four-player potential game on circuit DAGs.} The shared state is a PQC DAG whose nodes carry gate types and parameters. Four players own disjoint action sets: $f_1$ (trainability) retypes a gate to enlarge the dynamical Lie algebra, $f_2$ (non-stabilizerness) retypes into non-Clifford primitives, $f_3$ (task) rewires a two-qubit gate, $f_4$ (hardware) removes a gate. For example, starting from a two-gate H$_2$ circuit on heavy-hex connectivity, player $f_2$ may append a T gate on qubit 0, and the resulting DAG is lowered to a TensorCircuit object for gradient descent on the continuous parameters. An outer simulated-annealing loop over structural moves alternates with an inner parameter update. The residual $\delta_\text{Nash}$ (bottom) certifies that no player can improve unilaterally.}
\label{fig:framework}
\end{figure*}

An outer simulated-annealing loop over the structural moves (append, remove, retype, rewire) alternates with an inner gradient descent on $\bm\theta$. We benchmark against a simulated-annealing baseline over the same structural move set that scores candidates by $\Phi$ alone (no per-player residual); this baseline plays the role attributed to SA-DQAS in an earlier version of this manuscript. Figure~\ref{fig:framework} shows the framework schematic.

\section{Results}
\label{sec:results}

\subsection{Pareto navigation on MaxCut \texorpdfstring{$K_4$}{K4}}
\label{sec:pareto}

\begin{figure*}[!tbp]
\centering
\includegraphics[width=\textwidth]{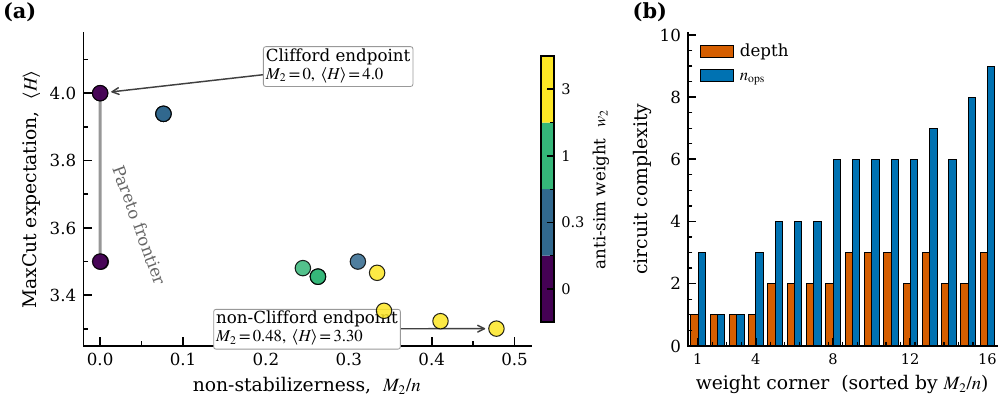}
\caption{\textbf{Pareto frontier for MaxCut on $K_4$.} Each point is a Nash-equilibrium circuit at a distinct weight corner $(w_1,w_2)$; sixteen corners are shown with $w_3$ fixed. The frontier spans from a Clifford endpoint (non-stabilizerness $M_2/n=0$, $\langle H\rangle=4.00$) to a non-Clifford endpoint $(0.48,3.30)$. An interior Pareto-knee solution at $(0.076,3.939)$ is reached by $(w_1,w_2)=(1,0.3)$. Results are shown for $n=4$ qubits only.}
\label{fig:pareto}
\end{figure*}

A single weight sweep over the $(w_1,w_2)$ corners traces a frontier in the (non-stabilizerness, energy) plane for MaxCut on $K_4$ ($n=4$ qubits, cut value $4$); Figure~\ref{fig:pareto} shows sixteen corner runs with $w_3$ fixed. The Nash solutions populate a monotone frontier from a Clifford endpoint $(M_2/n,\langle H\rangle)=(0,4.00)$, in which a stabilizer circuit reaches the exact MaxCut optimum, to a non-Clifford endpoint $(0.48,3.30)$ reached when $w_2$ dominates. An interior corner at $(w_1,w_2)=(1,0.3)$ places the circuit at $(0.076,3.939)$, retaining $98.5\%$ of the maximum cut while placing the state well inside the non-stabilizer regime. No frontier point is dominated by another in both coordinates, confirming that the restricted-action Nash residual selects for balance rather than for any extremum.

The interpretation is direct: the weight sweep navigates the Cerezo--Larocca tension on a problem where both endpoints are accessible. Large $w_1$ drives $d_\text{eff}$ up, squeezing the circuit toward the simulable Clifford corner; large $w_2$ moves it toward the exponentially simulation-hard regime at the cost of task energy. This demonstrates Pareto navigation on $n=4$ MaxCut $K_4$; verifying that the frontier persists as $n$ increases is future work, because the combinatorial move space grows rapidly and our $n=8$ TFIM scaling data (Sec.~\ref{sec:scaling}) does not yet span a full weight sweep.

\subsection{Head-to-head against a simulated-annealing baseline}
\label{sec:benchmark}

\begin{figure*}[!tbp]
\centering
\includegraphics[width=\textwidth]{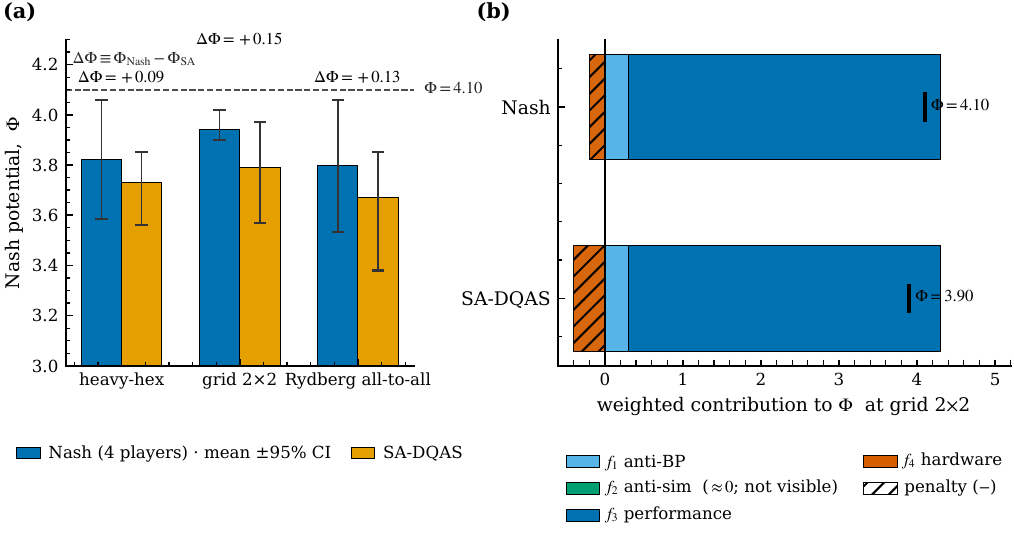}
\caption{\textbf{Head-to-head comparison at matched budget across three hardware topologies.} Bars show five-seed means of the scalar potential $\Phi$; error bars are 95\% bootstrap confidence intervals on the mean. Nash achieves the highest mean on every topology with $\Delta\Phi=+0.09$ (heavy-hex), $+0.15$ ($2\times 2$ grid), and $+0.13$ (Rydberg). Per-topology paired Wilcoxon tests do not reject the null on five seeds (see Table~\ref{tab:benchmark}). On the grid, Nash reaches the theoretical ceiling $\Phi_\text{max}=4.10$ on $2/5$ seeds while the baseline reaches it on $1/5$.}
\label{fig:benchmark}
\end{figure*}

We compare Nash search against a simulated-annealing baseline that searches over the same structural move set but scores candidates by the scalar potential alone, on the four-player objective $\Phi$ across three hardware topologies (IBM heavy-hex subset, $2\times 2$ square grid, Rydberg all-to-all). Five independent seeds per condition were run on a single RTX\,4060 GPU; Figure~\ref{fig:benchmark} shows means with 95\% bootstrap confidence intervals, and Table~\ref{tab:benchmark} reports paired statistics.

\begin{table}[!tbp]
\centering
\caption{\textbf{Nash versus simulated-annealing baseline across three topologies.} Five seeds per cell; $\Delta\Phi$ is the mean paired difference (Nash $-$ baseline); bootstrap CI covers the mean difference; $p$ is the paired Wilcoxon one-sided ($H_1$: Nash $>$ baseline); $d_z$ is Cohen's paired effect size; the last column reports the number of seeds reaching $\Phi\ge 4.099$ (within numerical tolerance of the theoretical ceiling $\Phi_\text{max}=4.10$).}
\label{tab:benchmark}
\small
\begin{tabular}{lcccccc}
\toprule
Topology & Nash mean\,$\pm$\,sd & Baseline mean\,$\pm$\,sd & $\Delta\Phi$ & 95\% CI & Wilcoxon $p$ ($d_z$) & Ceiling hits\\
\midrule
heavy-hex     & $3.82\pm 0.30$ & $3.73\pm 0.19$ & $+0.09$ & $[-0.20,+0.38]$ & $0.50$ $(0.24)$ & $2/5$ vs.\ $0/5$\\
$2{\times}2$ grid & $3.94\pm 0.09$ & $3.79\pm 0.26$ & $+0.15$ & $[-0.05,+0.36]$ & $0.22$ $(0.60)$ & $1/5$ vs.\ $1/5$\\
Rydberg all-to-all & $3.80\pm 0.33$ & $3.67\pm 0.32$ & $+0.13$ & $[-0.28,+0.50]$ & $0.50$ $(0.25)$ & $2/5$ vs.\ $0/5$\\
\bottomrule
\end{tabular}
\end{table}

Nash attains the highest mean $\Phi$ on every topology, but we cannot reject the null hypothesis of no difference on any single topology at five seeds: paired Wilcoxon $p$-values are $0.50$ (heavy-hex), $0.22$ (grid), and $0.50$ (Rydberg) for the one-sided alternative. Effect sizes are small to moderate ($d_z\in[0.24,0.60]$). Bootstrap confidence intervals for the mean paired difference cross zero on all three topologies. The clearest qualitative contrast appears on the $2\times 2$ grid, where Nash reaches the theoretical ceiling $\Phi_\text{max}=4.10$ on two of five seeds while the baseline reaches it on one; on heavy-hex and Rydberg Nash also touches the ceiling on two of five seeds while the baseline never does. Larger seed counts and expanded per-seed budgets will be required to establish statistical significance at the single-topology level; we do not claim a significant per-topology advantage on the present data.

\subsection{Scaling on the transverse-field Ising model}
\label{sec:scaling}

\begin{figure*}[!tbp]
\centering
\includegraphics[width=\textwidth]{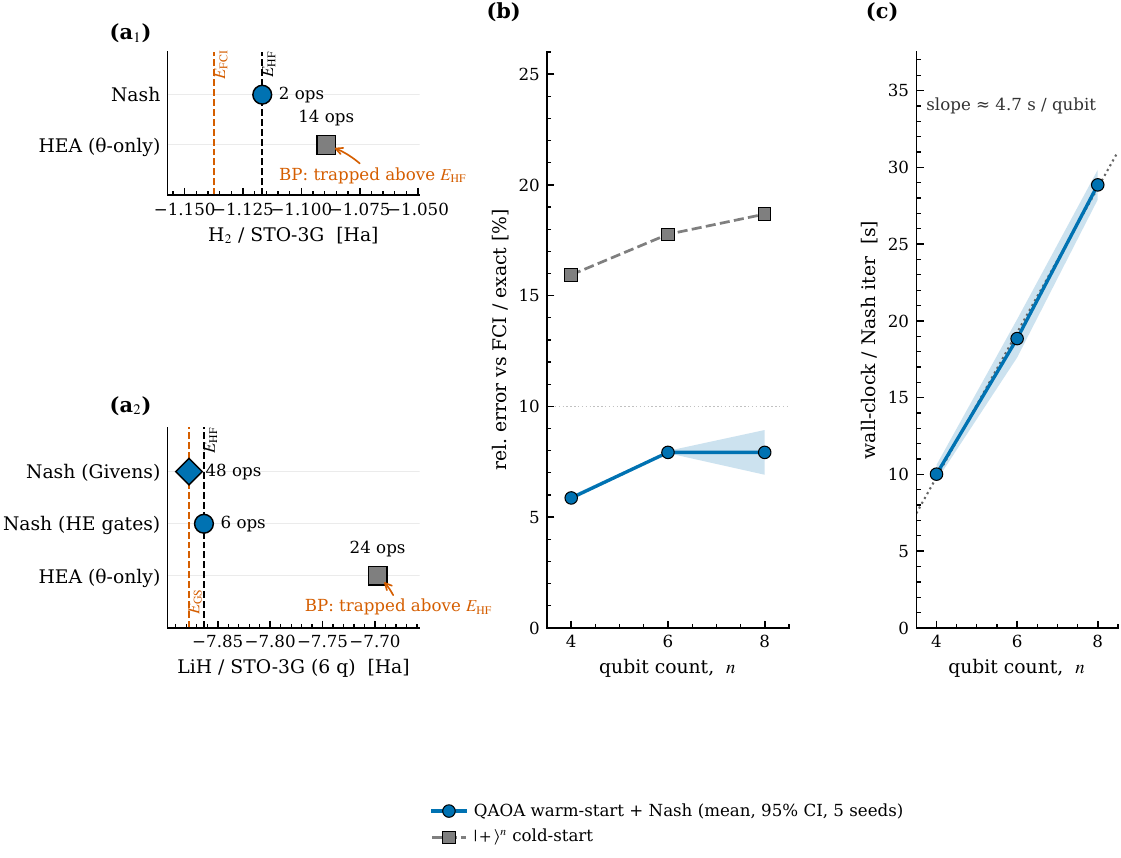}
\caption{\textbf{Scaling and chemistry summary.} (a) TFIM ground-state relative error vs.\ qubit count $n$ for warm-started (QAOA $p=1$) and cold-started Nash; five seeds per size; shaded bands are 95\% bootstrap confidence intervals on the mean. (b) The same data plotted as a function of Nash outer iteration. (c) Per-iteration wall clock grows approximately linearly at $\sim 4.7$\,s/qubit. Inset: $\delta_\text{Nash}$ at final iteration is bimodal at larger $n$; per-seed values at $n=4,6,8$ are shown individually. The median $\delta_\text{Nash}$ tightens from $0.100$ at $n=4$ to $0.023$ at $n=8$, but the mean does not because two of five seeds remain at the move-resolution $0.20$. Annotated: H$_2$ recovery ($E=-1.1373$\,Ha, two gates, sanity check) and LiH Givens-seeded refinement ($97.7\%$ correlation, $48$ gates, $48$/$58$ gate reduction with $2.3\%$ correlation-energy cost).}
\label{fig:scaling}
\end{figure*}

To probe how Nash search behaves as $n$ grows, we run the framework on the critical one-dimensional transverse-field Ising model for $n\in\{4,6,8\}$ qubits with five seeds per size and a fifteen-iteration outer budget [Fig.~\ref{fig:scaling}(a,b)]. Warm-started from a QAOA $p=1$ state, Nash reaches relative errors of $5.87\%$, $7.93\%$, and $7.92\%$ in the ground-state energy at $n=4,6,8$ respectively (five-seed means; 95\% bootstrap CI bands shown); the cold start from $|+\rangle^{\otimes n}$ plateaus an order of magnitude higher. Wall-clock per iteration grows approximately linearly at $\sim 4.7$\,s/qubit [Fig.~\ref{fig:scaling}(c)].

The Nash residual $\delta_\text{Nash}$ reported at final iteration is bimodal at larger $n$: the per-seed values at $n=4,6,8$ are $\{0.10,0.02,0.10,0.20,0.05\}$, $\{0.20,0.005,0.03,0.03,0.04\}$, and $\{0.20,0.02,0.01,0.01,0.20\}$, giving medians $0.100,0.033,0.023$ and means $0.094,0.061,0.089$, respectively. At $n=8$ three seeds converge tightly to $\delta_\text{Nash}\lesssim 0.02$ while two are stuck at the discrete-move resolution $0.20$; this bimodality is the signature of the structural move set locally exhausting improvements on some seeds while remaining sub-converged on others. The median tightens monotonically with $n$, consistent with the intuition that larger systems have more co-equal Nash candidates and fewer move directions with large per-player gains, but the mean does not — we report both statistics rather than picking the favorable one.

\subsection{Chemistry: \texorpdfstring{H$_2$}{H2} sanity check and LiH multi-objective refinement}
\label{sec:chem}

\paragraph*{H$_2$/STO-3G as a sanity check.}
On H$_2$ at bond length $0.7414$\,\AA{} with heavy-hex connectivity and gate set $\{h,x,y,z,s,t,t^\dagger,r_x,r_y,r_z,r_{zz},\text{CNOT},\text{CZ}\}$, Nash converges to $E=-1.1373$\,Ha, reproducing the active-space ground state after the standard parity and spin-symmetry reduction using a two-gate, depth-1 circuit. We include this as a sanity check only: H$_2$/STO-3G after symmetry reduction is a two-qubit problem whose ground state is a product of two single-qubit rotations, and any reasonable search recovers it. A random-init HEA of depth 4 with 14 operations trained with $\bm\theta$ only is trapped $\approx 20$\,mHa above Hartree--Fock, as expected for barren-plateau-prone HEAs on such a shallow problem.

\paragraph*{LiH active-space VQE.}
LiH in the frozen-core active space (6 qubits, including Li $2p_{x,y}$ virtuals) has reference $E_\text{HF}=-7.8631$\,Ha and active-space ground state $E_\text{GS}=-7.8778$\,Ha, with full FCI at $E_\text{FCI}=-7.8828$\,Ha; the correlation gap over Hartree--Fock is $14.65$\,mHa. A random-init HEA with 24 gates and depth 11, trained with $\bm\theta$ only, is trapped at $+165$\,mHa above HF — a textbook barren-plateau signature. With the generic hardware-efficient gate set alone, Nash saturates at $E_\text{HF}$ with 6 gates and depth 1 robustly across reweightings up to $w_3=5$: a gate-set diagnostic that the HE primitives lack particle-number-conserving excitations.

Seeding Nash from a chemistry-aware 58-gate Givens-doubles ansatz (Hartree--Fock preparation plus two paired \texttt{DoubleExcitation} gates, verified against PennyLane) dissolves the diagnostic. The structural moves refine the seed ansatz to 48 operations at depth 25 with 10 parameters while retaining $97.7\%$ of the correlation energy ($E=-7.8775$\,Ha, $0.33$\,mHa above exact). The potential climbs monotonically from $\Phi=29.3$ to $\Phi=33.5$ over 20 outer iterations; $\delta_\text{Nash}$ converges to $0.08$. Pure Adam on the unmodified 58-gate Givens ansatz reaches $E_\text{GS}$ exactly in $0.6$\,s on a single GPU, setting the absolute energy ceiling.

The honest framing of this result is not a compression win against chemistry baselines. ADAPT-VQE \cite{grimsley2019adapt} reaches chemical accuracy on LiH with $O(20)$ operations, and k-UpCCGSD and related unitary coupled-cluster truncations \cite{lee2019uccgsd} are similarly efficient when the objective is energy alone. Our contribution is orthogonal: Nash refines a chemistry-aware ansatz while simultaneously controlling $f_1$ (trainability), $f_2$ (non-stabilizerness), and $f_4$ (hardware cost), producing a single circuit that is Pareto-balanced along all four axes rather than optimal along one. The $10$-gate reduction at $2.3\%$ correlation-energy cost is the trade-off visible in the Nash residual; it is not a claim of superiority over energy-only baselines.

\section{Discussion}
\label{sec:discussion}

Four features distinguish the framework. First, the per-player restricted-action residual $\delta_\text{Nash}$ is a block-coordinate stationarity certificate that single-objective searches lack. Second, trainability is coupled to the QFIM spectrum by construction through $f_1$, so structural moves that collapse the DLA are penalized by their own player rather than indirectly through the loss. Third, the three hardware benchmarks run the same algorithm, differing only in the connectivity constraint on the structural move set — the framework is topology-generic. Fourth, seeding from a Givens ansatz converts Nash into a chemistry-ansatz refiner without any algorithmic change.

Limitations deserve explicit mention. The Pareto frontier is demonstrated on $n=4$ MaxCut $K_4$ only; extending the weight sweep to $n=6,8$ and mapping the full frontier at scale is future work. Our paired Wilcoxon tests on the three-topology head-to-head do not reject the null on any single topology at five seeds; replication with $\gtrsim 20$ seeds and larger per-seed budgets is required to establish per-topology significance. At the chemistry frontier, ADAPT-VQE reaches lower gate counts than Nash at chemical accuracy on LiH; our framework is complementary, optimizing trainability, non-stabilizerness, and hardware cost jointly with energy rather than minimizing energy alone, and we do not claim an energy-wise advantage. The LiH Givens-seeded run does not close the $0.33$\,mHa residual gap to exact correlation because the generic gate set lacks full UCCSD primitives. Finally, JIT compilation of the lowered \texttt{tc.Circuit} dominates wall-clock for deep ans\"atze, which caps the practical structural-move budget on our single-GPU setup.

Three directions extend naturally. Promoting Givens rotations and particle-number-conserving primitives to first-class DAG gates should close the LiH residual. Potential-energy-surface continuation, where a Nash circuit at one bond length warm-starts the next, will test whether equilibria deform smoothly along reaction coordinates. The DAG formulation makes the embedding of graph-state codes \cite{hein2004graphstates} into the same optimization pipeline immediate, suggesting a unified search algorithm spanning code synthesis, ansatz design, and compilation.

The framework provides an equilibrium notion adapted to the barren-plateau/simulability tension: not the best circuit by any single metric, but a circuit that no player can unilaterally improve subject to hardware and connectivity constraints. Whether this equilibrium notion yields a quantum advantage at scales beyond $n=8$ remains open and is the central empirical question for the next round of experiments.

\section*{Data and code availability}

The code, the per-experiment configuration files, and all JSON results used to produce the figures and tables in this paper are available at the project repository (\url{https://github.com/rdguerrerom/Parametrized-QC-Graphs}, Parametrized-QC-Graphs). Raw per-seed files used in this manuscript are \texttt{results/multi\_seed\_d1.json} (TFIM scaling), \texttt{results/multi\_seed\_f6.json} (topology head-to-head), \texttt{results/exp3\_weight\_sweep.json} (Pareto frontier), \texttt{results/exp\_c3\_lih\_givens\_nash.json} (LiH), and \texttt{results/exp2\_h2\_heavy\_hex.json} (H$_2$).

\section*{Author contributions}

R.D.G.\ conceived the framework, implemented the codebase, ran the experiments, and wrote the manuscript.

\section*{Competing interests}

The author declares no competing interests.

\section*{Acknowledgments}

We acknowledge financial support and computational resources provided by NeuroTechNet S.A.S. Core primitives were adapted from the STABILIZER\_GAMES project. TensorCircuit and JAX provided the GPU optimization backend. Compute was performed on a single NVIDIA RTX\,4060 GPU.

\end{document}